\documentclass[aps,prl,twocolumn,superscriptaddress]{revtex4-2}

\usepackage{afterpage}
\usepackage{enumitem}
\usepackage{eso-pic, graphicx}
\usepackage{natbib}
\usepackage[colorlinks=true , citecolor=blue,urlcolor=blue]{hyperref}
\usepackage{float}
\usepackage{url}
\usepackage{textcase}
\usepackage{bm}

\begin{document}

%\preprint{}

\title{Surface NMR using quantum sensors in diamond}


\newcommand{\affA}{Department of Chemistry, Chair of Physical Chemistry, Technical University of Munich,  Lichtenbergstra{\ss}e 4, 85748 Garching bei M{\"u}nchen, Germany}
\newcommand{\affB}{Walter Schottky Institute and Physics Department, Am Coulombwall 4, 85748 Garching bei M{\"u}nchen, Germany}
\newcommand{\affC}{Department of Chemistry, WACKER-Chair for Macromolecular Chemistry, Lichtenbergstra{\ss}e 4, 85748 Garching bei M{\"u}nchen, Germany}

\author{K. S. Liu}\affiliation{\affA}
\author{A. Henning}\affiliation{\affB}
\author{M. W. Heindl}\affiliation{\affA}\affiliation{\affB}
\author{R. D. Allert}\affiliation{\affA}
\author{J. D. Bartl}\affiliation{\affB}\affiliation{\affC}
\author{I. D. Sharp}\affiliation{\affB}
\author{R. Rizzato}\affiliation{\affA}
\author{D. B. Bucher*}\affiliation{\affA}
%\thanks{}


%\date{\today}

\begin{abstract}
Characterization of the molecular properties of surfaces under ambient or chemically reactive conditions is a fundamental scientific challenge. Moreover, many traditional analytical techniques used for probing surfaces often lack dynamic or molecular selectivity, which limits their applicability for mechanistic and kinetic studies under realistic chemical conditions. Nuclear magnetic resonance spectroscopy (NMR) is a widely used technique and would be ideal for probing interfaces due to the molecular information it provides noninvasively. However, it lacks the  sensitivity to probe the small number of spins at surfaces. Here, we use nitrogen vacancy (NV) centers in diamond as quantum sensors to optically detect nuclear magnetic resonance signals from chemically modified aluminum oxide surfaces, prepared with atomic layer deposition (ALD). With the surface NV-NMR technique, we are able to monitor in real-time the formation kinetics of a self assembled monolayer (SAM) based on phosphonate anchoring chemistry to the surface. This demonstrates the capability of quantum sensors as a new surface-sensitive tool with sub-monolayer sensitivity for \textit{in-situ} NMR analysis with the additional advantage of a strongly reduced technical complexity.
\end{abstract}

%\keywords{}

\maketitle

\section{Introduction}

Characterization of surface processes at the molecular level is important for understanding fundamental processes that are key to industrial catalysis, energy conversion, electronic circuits, targeted drug delivery, and biosensing \cite{somorjaiImpactSurfaceChemistry2011}. However, many analytical techniques used in surface science are inaccessible under ambient or chemically relevant conditions. Therefore, it remains challenging to perform chemical analysis under the conditions in which these processes occur \cite{salmeronSurfacesInterfacesAmbient2018, velasco-velezAtmosphericPressureXray2016}. Typical surface-sensitive methods, such as X-ray photoelectron spectroscopy (XPS), Auger electron spectroscopy and secondary ion mass spectroscopy (SIMS) can perform chemical analysis but require ultra-high vacuum conditions and expensive equipment  \cite{bolliESCAToolExploration2020}. To address this gap, great efforts have been devoted to extending XPS analysis to near ambient conditions \cite{salmeronSurfacesInterfacesAmbient2018}. Indeed, both near ambient pressure XPS and extended X-ray absorption fine structure (EXAFS) have greatly extended the applicability of these techniques for understanding reaction mechanisms at chemically active interfaces \cite{salmeronSurfacesInterfacesAmbient2018, wangSituXrayAbsorption2019}. However, to achieve high sensitivity and resolution  both methods require intense synchrotron radiation which limits their practical accessibility and increases their cost. Ultrafast spectroscopy techniques, such as sum frequency generation (SFG) and second harmonic generation (SHG) techniques, can perform analysis under ambient conditions but require technically demanding equipment, such as femtosecond lasers. Additionally, the extracted data are often difficult to interpret \cite{rosenfeldStructuralDynamicsCatalytic2011a}. Even with all these techniques available, molecular dynamics or chemical reaction kinetics at surfaces are still difficult to probe experimentally \cite{talapinFunctionalMaterialsDevices2020}.

Nuclear magnetic resonance (NMR) spectroscopy is one of the major tools for chemical and structural analysis in chemistry and material science. Solid-state NMR in particular \cite{rankinRecentDevelopmentsMAS2019} has advanced understanding of a range of systems including metal organic frameworks (MOFs) \cite{hoffmannSolidStateNMRSpectroscopy2012}, batteries \cite{pecherMaterialsMethodsNMR2017a}, and catalysts \cite{coperetActiveSitesSupported2017}. However, sensitivity remains a challenge for traditional NMR spectroscopy, making studies at surfaces difficult due to the limited numbers of nuclear spins. Even in highly porous materials with greater than 1000 m$^{2}$/g surface area, the concentration of NMR active nuclei of interest often remains low (e.g. 1 mmol of surface atoms/g) \cite{walderOneTwoDimensionalHighResolution2019a}, which requires long averaging times to obtain solid-state NMR spectra with reasonable signal-to-noise ratio (SNR). As a consequence, the technique is not broadly suitable for \textit{in-situ} kinetic studies of diverse materials over typical reaction time scales. Recently, surface enhanced NMR spectroscopy relying on hyperpolarization, such as dynamic nuclear polarization (\cite{rossiniDynamicNuclearPolarization2013, walderOneTwoDimensionalHighResolution2019a} or xenon based techniques \cite{haakeSurfaceEnhancedNMRUsing1997} gained interest. However, these methods are still limited to porous materials or require powders which can be difficult to prepare \cite{walderOneTwoDimensionalHighResolution2019a}.
 
 Here, we demonstrate the use of quantum sensors in diamond as a new surface-sensitive spectroscopy technique that works at ambient conditions and can probe planar interfaces on the microscopic length-scale with far greater sensitivity than conventional NMR in a totally non-invasive manner and at much lower costs and reduced technical complexity. The spectroscopic technique relies on the nitrogen vacancy (NV) point defect, consisting of a nitrogen impurity (N) and an adjacent vacancy (V) in the carbon lattice of diamond. These spin-1 defects allow for optical detection of magnetic resonance (ODMR) and have been established as highly sensitive nanoscale magnetic field sensors \cite{balasubramanianNanoscaleImagingMagnetometry2008, mazeNanoscaleMagneticSensing2008}. Shallow NV-centers are sensitive to magnetic fields from the Larmor precession of nuclei at the surface. This enables nanoscale NMR detection - even down to a single molecule \cite{lovchinskyNuclearMagneticResonance2016b} or spin \cite{mullerNuclearMagneticResonance2014, sushkovMagneticResonanceDetection2014}. The measurement volume of such NV sensors \cite{maminNanoscaleNuclearMagnetic2013, staudacherNuclearMagneticResonance2013} corresponds to a hemisphere whose radius is roughly their depth below the surface in the diamond lattice (e.g. 5-10 nm). At this small length scale, the thermal polarization of the nuclear spins can be neglected, since spin noise dominates for a small number of spins \cite{kehayiasSolutionNuclearMagnetic2017, merilesImagingMesoscopicNuclear2010}. For that reason the NMR signal strength is independent of the magnetic field B$_0$, reducing experimental complexity and costs, which makes the technique accessible to a broader community \cite{bucherQuantumDiamondSpectrometer2019b}. In most nanoscale NV-NMR experiments the sample (typically viscous oil) is placed directly on the diamond surface. In this work, we demonstrate that NV-centers not only can probe diamond surfaces, but can be used as a general tool to probe chemistry on a variety of surfaces. We used atomic layer deposition (ALD), a technology that can be applied to synthesize a wide variety of films with high thickness precision, to coat the diamond with amorphous aluminum oxide (Al$_2$O$_3$). Al$_2$O$_3$ provides an exemplary surface of high technical relevance in optoelectronic applications and  acts as structural support in a variety of catalytic processes \cite{ivanovaAluminumOxideSystems2012}. In a proof-of-concept study for this new surface-sensitive spectroscopic technique, we probe the chemical modification of the Al$_2$O$_3$ surface with phosphonate anchoring during the formation of a self-assembling monolayer (SAM) \cite{zhaoRecentDevelopmentPhosphonic2017}. We demonstrate that this technique offers the non-invasive quality of NMR with the advantages of sub-monolayer sensitivity without the cost or technical demands of other state-of-the-art surface analysis techniques.

\section{Results}

\begin{figure*}[t]
  \includegraphics[width=0.75\textwidth]{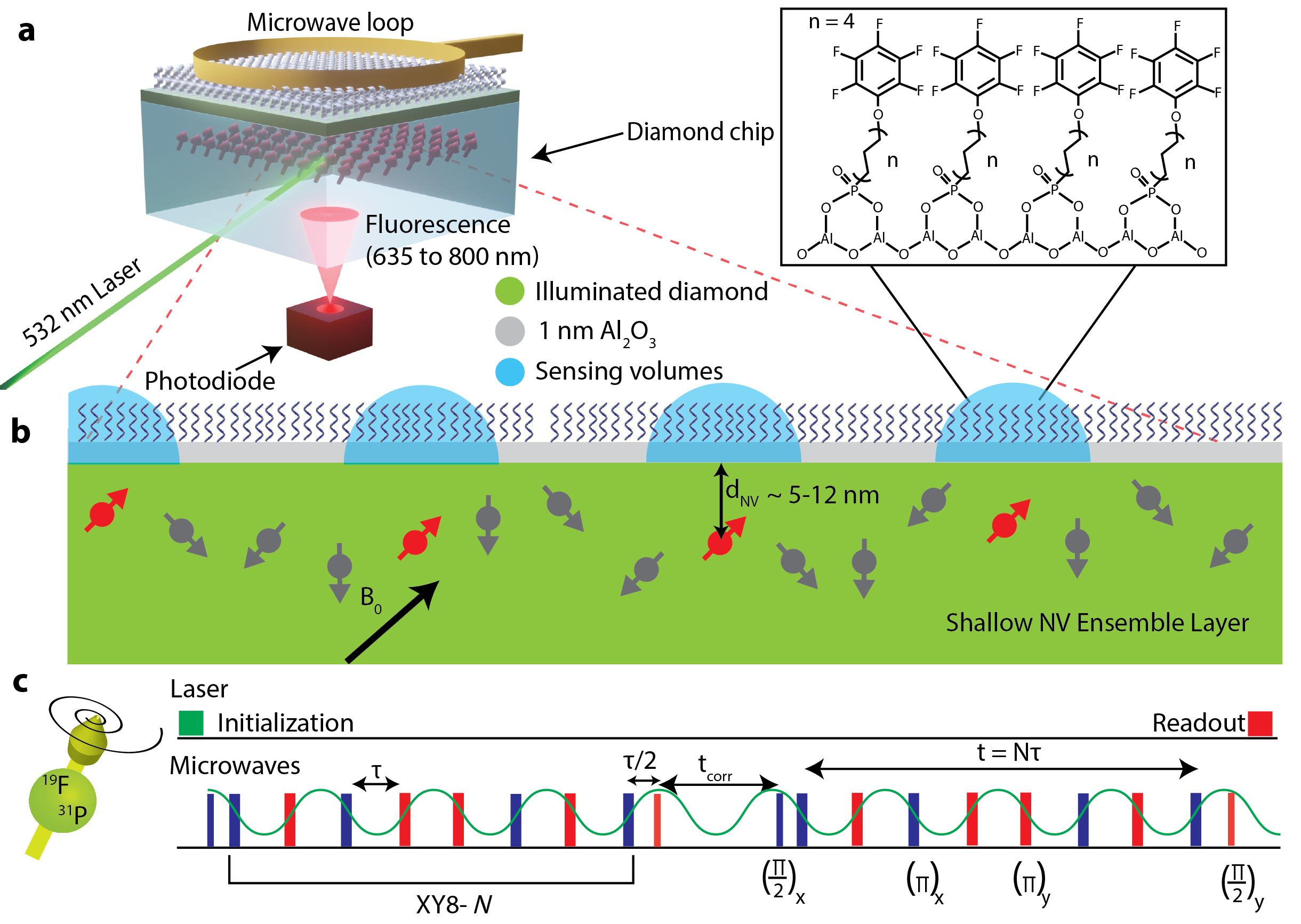}
  \caption{\textbf{Surface NV-NMR spectroscopy on a functionalized Al$_2$O$_3$  surface.} a) Schematic of the experiment. Near-surface NV-centers in a 2 mm x 2 mm x 0.5 mm diamond chip are excited with 532 nm laser in a total internal reflection geometry. The resulting spin-dependent photoluminescence from the  NV defects is detected with a photodiode. The microwave pulses for quantum control of the spin state of the defects are delivered through a small wire loop. b) NV-centers aligned with the magnetic field have sensing volumes with a radius determined by their distance to the surface, which is between 5-12 nm in our case. Inset: Schematic of organic monolayer formed from 12-pentafluorophenoxydodecylphosphonic acid (PFPDPA) on 1 nm Al$_2$O$_3$ deposited on the diamond surface by ALD. c) Correlation spectroscopy pulse sequence. Two blocks of dynamic decoupling XY8-\textit{N} sequences are correlated by sweeping the time between them (\textit{t$_{corr}$}). The time spacing $\tau$ between the $\pi$ pulses is set to half the period of the Larmor frequency of the nuclear spin being sensed. The NV spin state is initialized with a 532 nm laser pulse and PL detection with a photodiode occurs after the microwave pulse sequence.}
 \label{fig1}	
\end{figure*}

\begin{figure*}[t]
  \includegraphics[width=0.8\textwidth]{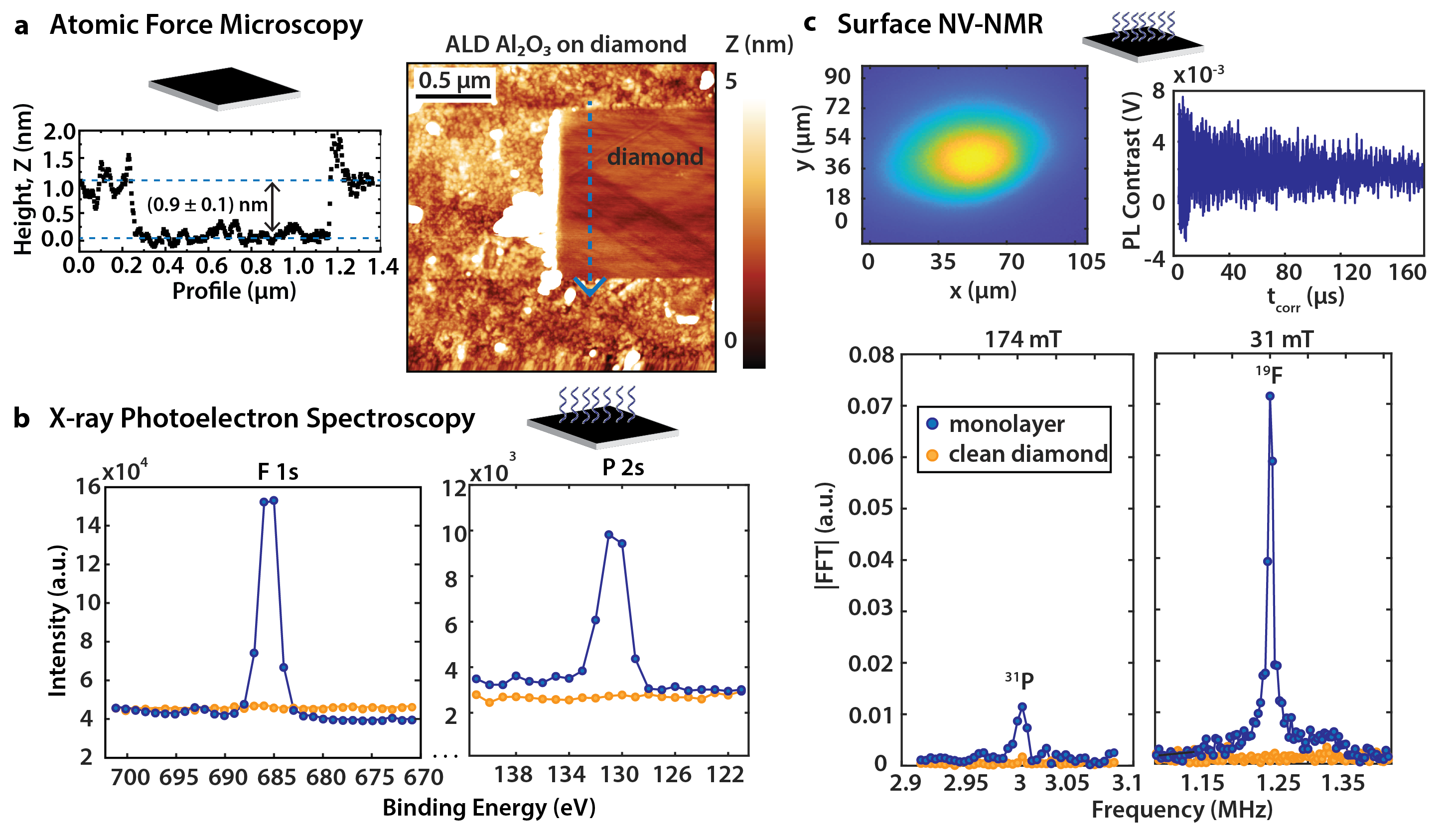}
  \caption{\textbf{Surface NV-NMR and validation with complimentary analytical surface techniques}. a) Diamond coated with a Al$_2$O$_3$ layer. The thickness of the ALD deposited Al$_2$O$_3$ layer was determined by AFM scratching measurements. The height profile along the segment indicated by a blue arrow was used to determine the Al$_2$O$_3$ film thickness of 0.9 nm $\pm$ 0.1 nm. b) Functionalized Al$_2$O$_3$ surface on diamond. The presence of PFPDPA molecules on the surface is confirmed with XPS by the appearance of F 1s and P 2s peaks (blue), which are absent on the clean diamond (yellow). c) Surface NV-NMR spectroscopy. Top: Image of the laser spot ($\sim$ 4000 $\mu$m$^2$) on the diamond and time domain correlation signal of $^{19}$F. Bottom: Surface NV-NMR spectrum of $^{31}$P detected from the monolayer measured at 174 mT and $^{19}$F nuclei detected at 31 mT. The clean diamond reference is shown in yellow.}
 \label{fig2}	
\end{figure*}

\begin{figure}[t]
  \includegraphics[width=0.49\textwidth]{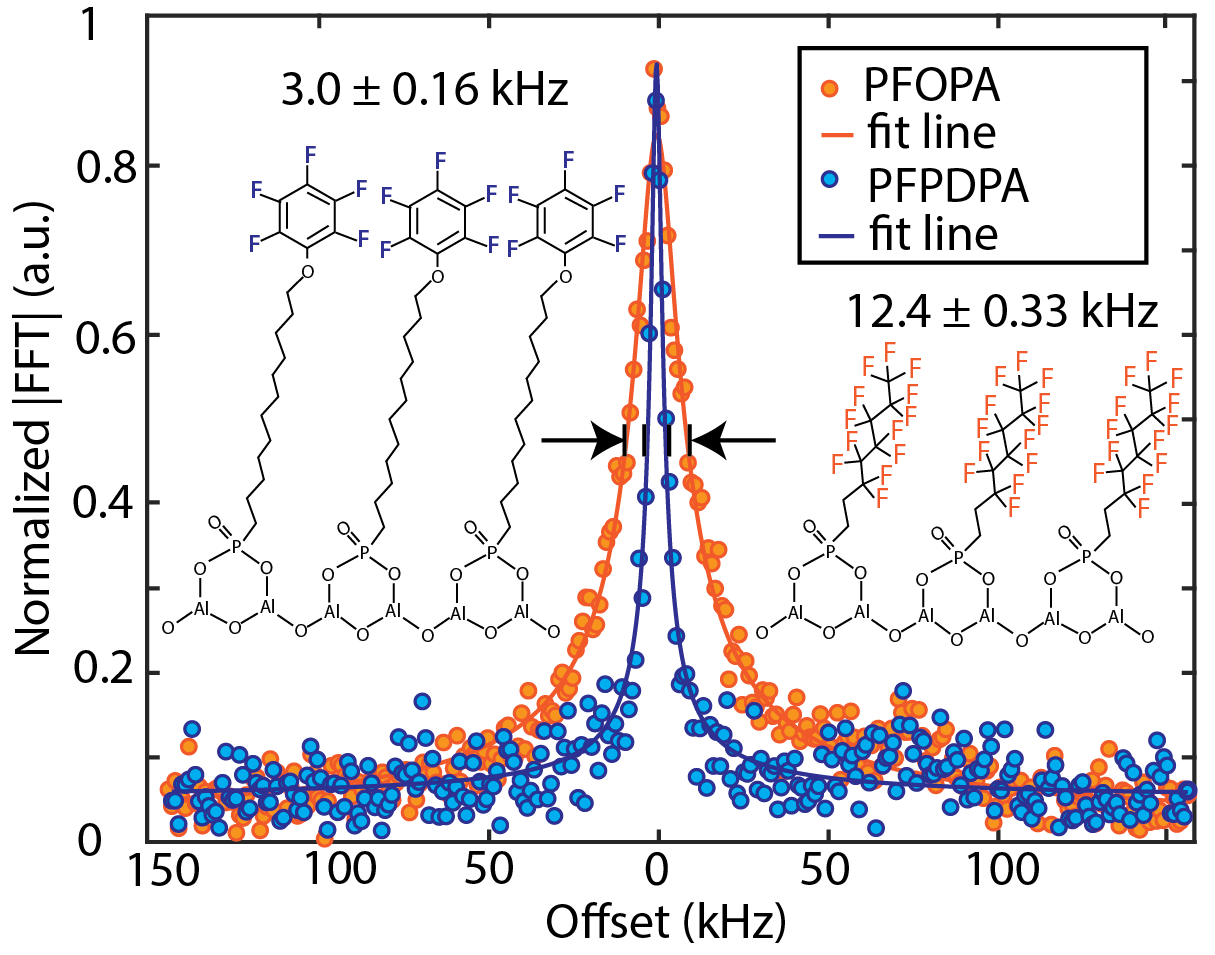}
  \caption{\textbf{Influence of molecular structure on $^{19}$F resonance linewidth.} The $^{19}$F linewidth of a monolayer made from PFOPA is $\sim$ 4 times broader than that of PFPDPA. This is likely caused by local dynamics of the fluorinated phenolic moeity which leads to line narrowing of the NMR signal.}
 \label{fig3}	
\end{figure}

\begin{figure*}[t]
  \includegraphics[width=0.75\textwidth]{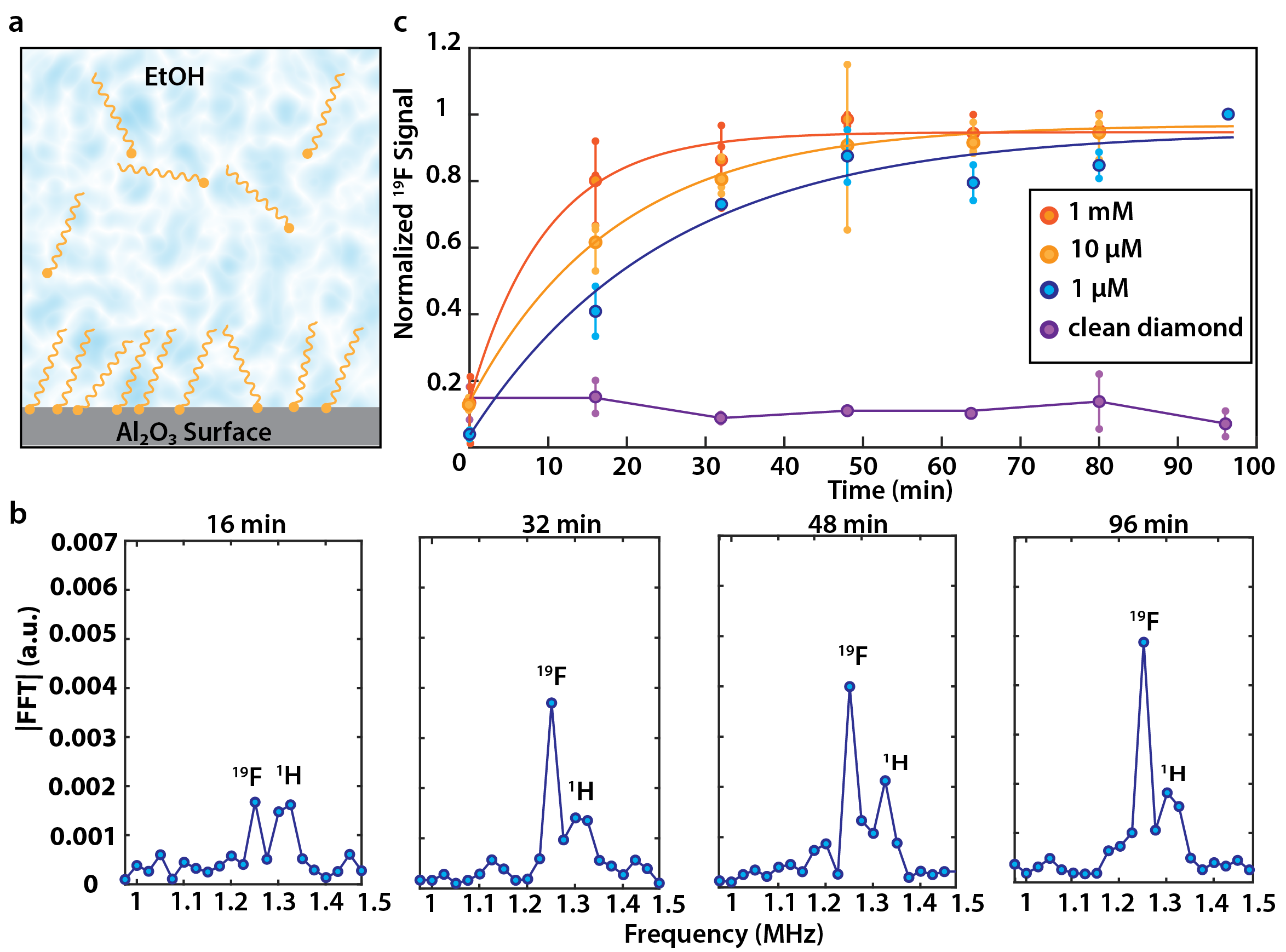}
  \caption{\textbf{Probing surface chemistry \textit{in-situ}.} a) Schematic of monolayer formation on Al$_2$O$_3$ in phosphonic acid (PA) solution. b) Individual spectra of $^{19}$F showing the time evolution for a 1$\mu$M solution. The peak around 1.35 MHz can be assigned to protons ($^{1}$H) which are known to be ubiquitous on or within the diamond and are not characteristic of the surface chemistry. c) Real-time monitoring of the $^{19}$F NMR signal amplitude growth in solution for three different PA concentrations shows a decrease in formation rate for lower concentrations. Small markers are measured data points of repeated experiments connected with a vertical line, averaged data are shown with the large points. Data points are fit with a single exponential. Background signal of a clean, non - Al$_2$O$_3$ coated diamond in 10 $\mu$M PA solution shows no $^{19}$F signal. Note that due to the absolute value of the Fourier transform, the noise floor is always positive.}
 \label{fig4}	
\end{figure*}

\textbf{Surface NV-NMR.} The surface NV-NMR technique is performed on a diamond chip in which $^{15}$N was implanted with a particle fluence of 2x10$^{12}$/cm$^{2}$ and with an energy of 2.5 keV (Fig. \ref{fig1} a) (see Supplementary Note 1 and 2). This results in a distribution of near-surface NV-centers $\sim$ 5-12 nm below the surface \cite{phamNMRTechniqueDetermining2016}. For these implant conditions, we estimate an NV density of $\sim$ 50-100 NVs/$\mu$m$^{2}$ \cite{ziemQuantitativeNanoscaleMRI2019, devienceNanoscaleNMRSpectroscopy2015, healeyHyperpolarisationExternalNuclear2021}, corresponding to 2-4x10$^5$ NV-centers for the $\sim$ 4000 $\mu$m$^2$ spot used in our experiments as shown in Fig. \ref{fig2} c (see Supplementary Note 3). These defects can have four different orientations in the tetrahedral diamond lattice; so one in four is aligned with the external magnetic field B$_0$. Consequently, an effective ensemble of $\sim$ 0.5-1x10$^5$ NV-centers allows for random NMR sampling of the diamond surface within the laser spot. The NMR detection volume of each NV center is determined by its depth, indicated schematically as blue hemispheres in Fig. \ref{fig1} b. 

The quantum sensing scheme for the detection of NMR signals with NV-centers in diamond has been described in detail before \cite{ bucherQuantumDiamondSpectrometer2019b, staudacherProbingMolecularDynamics2015, kehayiasSolutionNuclearMagnetic2017, laraouiHighresolutionCorrelationSpectroscopy2013}. In brief, the electronic ground state of the NV center is a spin triplet with the Zeeman states m$_{s}$ = 0 and $\pm$ 1, which are separated by approximately 2.87 GHz at zero magnetic field. The degenerate m$_{s}$ = $\pm$ 1 states are typically split by a static external magnetic field and transitions between the Zeeman split states can be addressed by microwave fields. The spin state of the NV center can be initialized in the m$_{s}$ = 0 state with laser excitation at a wavelength of 532 nm and optically read out due to spin-dependent photoluminescence (PL), which is weaker for the  m$_{s}$ = $\pm$ 1 states compared to the m$_{s}$ = 0 state. After optical excitation nearly all the NV spins are in the m$_{s}$ = 0 state. Subsequently, transitions between the m$_{s}$ = 0 and m$_{s}$ = $\pm$ 1 can be coherently controlled with microwave pulses. For nuclear spin detection, dynamic decoupling sequences (such as XY8-\textit{N}) are used. These sequences are sensitive to frequencies corresponding to 1/(2$\tau$), where $\tau$ is the spacing between the $\pi$ pulses (Fig. \ref{fig1} c). Sweeping the time \textit{\textit{t$_{corr}$}} between two XY8-\textit{N} sequences correlates oscillating magnetic signals, such as spin noise, from nanoscale nuclear spin ensembles. The detected spin noise appears as oscillations in the PL readout as a function of \textit{\textit{t$_{corr}$}} and resembles the free induction decay in traditional NMR spectroscopy. However, this nanoscale NMR spectroscopy can be performed without expensive magnets due to the detection of spin noise. Importantly, there is no need to excite the nuclear spins with radiofrequency pulses or to wait for nuclear spin-lattice relaxation.

\textbf{Characterization of the Al$_2$O$_3$ layer.} In previous work, NMR signals from samples directly on the diamond surface have been detected \cite{kehayiasSolutionNuclearMagnetic2017, devienceNanoscaleNMRSpectroscopy2015}. Here, the goal is to probe non-diamond surfaces and interfaces. This requires the preparation of a material of interest on top of the diamond substrate. In our proof-of-concept study we use an Al$_2$O$_3$ film prepared by ALD \cite{henning2021aluminum}, whose surface modification with organophosphonate chemistry shall be investigated through the surface NV-NMR technique. First, we optimized the thickness of the ALD layer by keeping it as thin as possible allowing the NV-centers to sense the surface modification, while also ensuring that it was thick enough to create a closed film onto which a dense molecular monolayer could be bound. The formation of the fluorinated monolayer on the Al$_2$O$_3$ surface increases the surface hydrophobicity, which can be investigated with static water contact angle (SWCA) measurements. The minimal ALD layer thickness required to facilitate the formation of a dense SAM was therefore determined using SWCA measurements as a function of Al$_2$O$_3$ thickness from 0.5 nm to 3 nm. Saturation of the SWCA signal appears for ALD layers of 1 nm and beyond (see Supplementary Note 4), which indicated that we reached the minimal thickness required for the organic monolayer to be fully formed. The hardness of the diamond chip allows for a scratching experiment to corroborate the thickness of the Al$_2$O$_3$ layer on the diamond using atomic force microscopy (AFM) (Fig. \ref{fig2} a). Removing the Al$_2$O$_3$ with the AFM tip revealed two different surfaces - the Al$_2$O$_3$ with a root-mean-square (RMS) roughness of 0.71 nm and the underlying diamond with a roughness of 0.25 nm. A vertical cut showed a step height of 0.9 nm $\pm$ 0.1 nm, confirming the thickness expected for the number of ALD cycles \cite{puurunenSurfaceChemistryAtomic2005}. Second, we ensure that the material preserves the NV center properties by quantifying the coherence times before and after depositing the Al$_2$O$_3$ film on the diamond. We observe a small reduction in the spin-lattice relaxation T$_1$ ($\leq$ 10 percent) and spin-spin relaxation T$_2$ ($\approx$ 25 percent) times (Table 1), with only a minor influence on the sensitivity. 

\begin{table}
\caption{Influence of Al$_2$O$_3$ on NV center relaxation properties}
\begin{tabular}{ |c|c|c| } 
\hline
 & T$_2$ ($\mu$s) & T$_1$ (ms)\\ 
\hline
Clean diamond  \linebreak (N = 5) & 5.59	$\pm$ 0.13 & 0.80 $\pm$	0.28 \\
\hline
After ALD  \linebreak (N = 8) & 4.19 $\pm$ 0.24 & 0.73 $\pm$ 0.30 \\ 
\hline
\end{tabular}
\end{table}

\textbf{Chemical characterization of the functionalized Al$_2$O$_3$ support.} Following thickness optimization, we analyzed the chemical composition of the functionalized Al$_2$O$_3$ layer. We selected a C$_{12}$ chain phosphonic acid (PA) terminated with a fluorinated phenolic ring (12-pentafluorophenoxydodecylphosphonic acid (PFPDPA)) for chemical modification of the Al$_2$O$_3$ surface. Fig. \ref{fig1} a illustrates the bridged bi-dentate binding motif. We note, however, that multiple binding modes might be present on the surface \cite{pujariCovalentSurfaceModification2014}. These flourinated monolayers can be easily prepared by soaking the Al$_2$O$_3$/diamond in the PA solution. The functionalization occurs via the binding of phosphonic acid groups to the hydroxy groups of the Al$_2$O$_3$ surface \cite{pujariCovalentSurfaceModification2014}. X-ray photoelectron spectroscopy (XPS) confirmed the presence of the F 1s peak and P 2s peaks from the monolayer (Fig. \ref{fig2} b). Corresponding spectra from the bare substrate prior to functionalization did not contain and detectable F or P. Thus, the phosphorous and fluorine peaks originate from the PFPDPA self-assembled monolayer bound to the Al$_2$O$_3$ surface.

The surface NV-NMR technique is capable of providing chemical information from the SAM-functionalized  Al$_2$O$_3$ surface with high sensitivity, much like XPS but under ambient conditions. The $^{19}$F correlation spectroscopy data provides a time domain NMR signal (Fig. \ref{fig2} c), clearly showing a oscillation at the $^{19}$F Larmor frequency that decays. The Fourier transform of these data results in the $^{19}$F NMR spectrum with a resonance at 1.247 MHz, which agrees with the theoretical Larmor frequency at 31 mT.  Similarly, we can detect the $^{31}$P signal from the monolayer, which results in a peak at 3 MHz at 174 mT (see Supplementary Note 5). The signal is weaker since the number of spins per molecule and the gyromagnetic ratio are lower, compared to $^{19}$F. Both signals were taken from the same monolayer and laser spot.

We note that other NMR active nuclei are present in this system, most notably $^{27}$Al and $^{1}$H. However, the strong  $^{13}$C signal naturally occurring in the diamond overlaps with the $^{27}$Al resonance, precluding detection in the current experiment. This can be overcome in the future by using an isotopically enriched $^{12}$C diamond. The $^{1}$H signal is ubiquitous which is also observed on a clean diamond and cannot be unambiguously attributed to a characteristic of our system. Therefore, it has been excluded from the present analysis.
%We can estimate the number of spins that contribute to our signal with our contact angle using Cassie's equation as well as surface NV-NMR with a sample of known density in comparison to our monolayer (see SI section about the contact angle \ref{contact_angle} and section about calibration \ref{Fomblin_calibration} for more details about the calculations). Both estimates are consistent with each other. In our elliptical laser spot of $\sim$ 300 $\mu$m$^2$ we estimate a total of $\sim$ 5 x 10$^6$ $^{19}$F spins and 1 x 10$^6$ $^{31}$P spins. This is a sensitivity unmatched by conventional NMR, which requires spins on the order of 10$^{12-15}$ and allows us to monitor a chemical reaction occurring at the surface.

\textbf{Characterization of molecular dynamics at the surface.} 
From our measurement of the PFPDPA monolayer a resonance linewidth of $\sim$ 3 kHz was observed, which is narrower than what has been commonly seen in previous NV-NMR experiments of solid samples \cite{aslamNanoscaleNuclearMagnetic2017, devienceNanoscaleNMRSpectroscopy2015}. For that reason we performed a second set of experiments in which the Al$_2$O$_3$ surface was functionalized with a shorter, but perfluorinated PA molecule (1H,1H,2H,2H-perfluoroctanephosphonic acid (PFOPA)) (Fig. \ref{fig3}). The resonance linewidth is much broader ($\sim$ 12 kHz) than for the case of the monolayer made from PFPDPA. In the solid state, the linewidth is typically limited by dipolar broadening, which can be minimized by local molecular dynamics such as rotations. The fluorinated phenolic moeity attached to a long carbon chain is more mobile than the $^{19}$F nuclei in the perfluorinated chain of PFOPA, which likely reduces the linewidth \cite{gaborieauChainDynamicsPoly2007, purschChainOrderMobility1996, vugmeysterStaticSolidstate2H2017}.

\textbf{Monitoring surface chemistry in real-time.} In contrast to other surface sensitive techniques, surface NV-NMR allows for measurements under chemically relevant conditions, for example at the solid-liquid interface. In the present case, this enables the observation of the binding or the phosphonate anchoring group to the Al$_2$O$_3$ support at the solid/liquid interface, which is depicted in Fig. \ref{fig4} a. In particular, the chemical reaction kinetics were directly detected by addition of PFPDPA solution onto a freshly prepared Al$_2$O$_3$ layer on a diamond and measurement of the surface NV-NMR signal as a function of time. \ref{fig4} b shows individual surface NV-NMR spectra at different times after adding the PFPDPA solution. The $^{19}$F resonance signal grows on a time scale of tens of minutes and plateaus after approximately half an hour. The broad resonance at 1.325 MHz next to the $^{19}$F signal originates from $^{1}$H, which is present from the beginning and is not characteristic of the surface chemistry. We repeated the experiment using solutions of different concentrations of the PFPDPA, which showed an increase in the monolayer formation rate at higher concentration (Fig. \ref{fig4} c). The kinetics of the $^{19}$F signal can be modelled with an exponential growth function (see Supplementary Note 6) \cite{dietrichMolecularMechanismsSolventControlled2017, koutsioubasFormationAlkanephosphonicAcid2009, meltzerMolecularStructureOctadecylphosphonic2018}. In a reference experiment, we repeated the experiment with a clean diamond without a Al$_2$O$_3$ layer. In this case we did not see any signal at the expected $^{19}$F resonance which indicates that the monolayer is inefficiently formed on diamond. 

\section{Conclusions}
The surface NV-NMR technique has been successfully demonstrated to enable the detection of microscopic NMR signals down to sub-monolayer coverages, including with real-time formation of molecular monolayer assembly on Al$_2$O$_3$. The use of a diamond-based sensor that is chemically inert and can withstand high temperatures and high pressures is advantageous especially, for chemical applications and catalysis. It will not only bridge the “pressure gap” in surface science but has the resilience to probe chemical reactions even under harsh conditions \textit{in-situ} \cite{velasco-velezAtmosphericPressureXray2016}. Yet, the technique is still in its infancy and has many possible avenues for improvement. The main challenge is the lack of molecular structural information such as chemical shifts in comparison to traditional NMR spectroscopy. The main reason for the broad (a few kHz) resonances is the dipolar broadening in the solid state. Magic angle spinning, typically used in solid state NMR, may be one way to reduce the linewidth but comes with significant engineering challenges \cite{woodQuantumMeasurementRapidly2018}. Alternatives are decoupling pulse sequences, which reduce sensitivity and cannot fully recover high resolution spectra \cite{aslamNanoscaleNuclearMagnetic2017}. Another approach is the use of quadrupolar nuclei such as ($^{14}$N, $^{2}$D, etc.) which have been shown to convey detailed structural information despite their broad lines making them ideal targets for surface NV-NMR \cite{vugmeysterStaticSolidstate2H2017}. 

Although the sensitivity did not limit this study, future analysis of sparsely modified surfaces or the detection of nuclei with low gyromagnetic ratios would benefit from further improvement. Advanced NV generation techniques \cite{luhmannCoulombdrivenSingleDefect2019, favarodeoliveiraTailoringSpinDefects2017} or the growth of preferentially orientated NV-centers \cite{osterkampEngineeringPreferentiallyalignedNitrogenvacancy2019} can improve the sensitivity significantly. Additionally, an improved readout scheme of the NV quantum state has been shown to increase the sensitivity for single NV-centers by over an order of magnitude \cite{lovchinskyNuclearMagneticResonance2016b}. Finally, it should be noted that NV-centers are strongly influenced by magnetic noise. Consequently, para-magnetic materials on the diamond surface will deteriorate the NV coherence times and preclude surface NV-NMR studies of this type. However, other quantum sensing schemes are available to study these types of materials \cite{steinertMagneticSpinImaging2013}. 

In summary, the results from this study introduce a new surface-sensitive technique apt for probing the chemical composition of surfaces and chemical reaction kinetics at interfaces. Using the ability of NV-centers to detect NMR signals selectively from nanoscale layers with well-established and versatile ALD techniques, make this a powerful and broadly applicable. As a proof-of-concept, we demonstrate the detection of NMR signals from a phosphonate functionalized Al$_2$O$_3$ surface. In contrast to other surface techniques, sub-monolayer signals can be detected under ambient conditions. This enables us to monitor the chemical formation of the monolayer in real-time in solution at the solid-liquid interface on the microscopic level. We believe that this will facilitate further understanding of a variety of surface phenomena. Not only does this method offer non-invasive benefit of NMR spectroscopy, its functionality under chemically relevant conditions and the low technical complexity make it a practical and sensitive technique for advanced studies in important areas of catalysis, materials science, biological sensing or 2D materials research.

\section{Methods}

\textbf{Diamond preparation.}
An electronic grade diamond (1.1$\%$ $^{13}$C abundance, Element Six) was implanted with $^{15}$N at an energy of 2.5 keV with an off axis tilt of 7 degrees and with a fluence of 2x10$^{12}$/cm$^{2}$ by Innovion. Then, the implanted sample was annealed under vacuum in a home built oven for over 32 hours (see Fig. \ref{temperature_profile}) with a Tectra BORALECTRIC\textsuperscript{\textregistered} sample heater. The diamond was cleaned before every Al$_2$O$_3$ deposition with a tri-acid cleaning protocol involving equal parts boiling sulfuric, nitric, and perchloric acid according to Brown \textit{et al.} \cite{BROWN201940}.

\textbf{Atomic layer deposition.}
 ALD was performed using a Veeco Fiji G2 system. Prior to deposition, the diamond surface was cleaned \textit{in-situ} by 5x (0.15)s cycles of ozone, which was generated via electrical discharge in O$_2$. For deposition of Al$_2$O$_3$ thin films, the ozone-treated diamond substrates were sequentially exposed to trimethyl aluminum (98-1955, STREM Chemicals) followed by H$_2$O at 200$^{\circ}$ C cyclically. Each ALD cycle followed the sequence: TMA pulse/Ar purge/H$_2$O pulse/Ar purge. This was repeated for 10 cycles to achieve a final thickness of approximately 1 nm (nominal growth per cycle of 0.1 nm/cycle). To achieve a reproducible surface hydroxyl surface termination suitable for monolayer assembly by phosphonate chemistry, the sample was exposed to a remote oxygen plasma within the ALD system. In particular, the RF (13.64 MHz) inductively coupled plasma (ICP) source of the Fiji G2 system was operated at 300 W for a total of 1 min exposure. For repeated use of diamond substrates, the Al$_2$O$_3$ was removed by soaking overnight in a 5$\%$ NaOH solution (see Supplementary Note 7) before the next ALD deposition.
 
 \textbf{Phosphonic acid surface functionalization}. For the fully formed monolayer shown in Fig. \ref{fig2} and \ref{fig3}, the following procedure was used. After the ALD process, the diamond is immersed in a 10 mM solution of either 12-pentafluorophenoxydodecylphosphonic acid (CAS-Number 1049677-16-8, Sigma-Aldrich) or 1H,1H,2H,2H-perfluorooctanephosphonic acid (CAS-Number 252237-40-4, Sigma-Aldrich) in ethanol for 2 days to allow the monolayer formation to equilibrate. Then the diamond is sonicated for 5 minutes in ethanol to remove all physiadsorbed molecules, after which the sample was dried with flowing nitrogen. 

\textbf{Surface NV-NMR.}
 The experiment is based on a modified version of the setup described in Bucher \textit{et al.} \cite{bucherQuantumDiamondSpectrometer2019b} (see details in Supplementary Note 8). Correlation spectroscopy was performed using XY8-\textit{4} blocks (a total of 32 $\pi$ pulses) with \textit{\textit{t$_{corr}$}} swept starting from 2 $\mu$s to obtain the spectra. For $^{19}$F detection \textit{\textit{t$_{corr}$}} was swept until 160 $\mu$s in 801 points. The time domain data was then Fourier transformed and the absolute value plotted using MATLAB. Each spectrum shown in Fig. \ref{fig2} c is zero-filled with 801 points. For the $^{19}$F signal shown in Fig. \ref{fig2} c, only 1 average was necessary to obtain a SNR of 95, as calculated by dividing the signal value by the standard deviation of the noise floor within a region without signal. For $^{31}$P detection, \textit{t$_{corr}$} was swept until 80 $\mu$s in 801 points. The $^{31}$P signal was averaged 2 times for an SNR of 15. The $^{19}$F and $^{31}$P NMR signals were obtained in 25 minutes and 32 minutes, respectively. For linewidth measurements as shown in Fig. \ref{fig3}, each monolayer was measured to 2501 points and \textit{t$_{corr}$} swept to 0.5 ms. After zero-filling to 5001 points, the linewidth of the resonance was fit with a modified Lorentzian model \cite{bucherHyperpolarizationEnhancedNMRSpectroscopy2020}. 
 
 \textbf{\textit{In-situ} kinetics of monolayer formation.}
 The tri-acid cleaned diamond was first coated with 1 nm of Al$_2$O$_3$ and activated with oxygen plasma, as described above. Subsequently, the diamond was glued down to a water-tight liquid sample holder made from a 30 mm cage plate (CP4S, Thorlabs) with a thin round cover slide (100493-678, VWR) glued to the bottom and the top fitted with a threaded lens tube (SM30L03, Thorlabs). The cage plate was then mounted to the surface NV-NMR experiment (see Supplementary Note 8). $^{19}$F is detected with 24 time traces, which were continuously acquired by sweeping \textit{t$_{corr}$} to 40 $\mu$s with 201 points. Each data point in Fig. \ref{fig4} b was an average of 4 time traces, which were then Fourier transformed and the $^{19}$F signal amplitude plotted, resulting in 6 points for the kinetic dataset. Each of these experiments are repeated 3 times (1 mM, 10 $\mu$ M) and 2 times (1 $\mu$M) and the signal amplitudes are averaged and then fit. Each time trace is normalized to 1 for the time point at 96 minutes. The first point at t = 0 minutes is set to the mean of the noise floor of the Fourier transformed spectra. The background signal is the value at the $^{19}$F frequency within spectra obtained with a clean diamond measured in the intermediate concentration 10 $\mu$M solution. As there is no $^{19}$F, the background data is normalized to the final $^{19}$F signal of a 10 $\mu$M growth kinetics data set.
%\begin{acknowledgments}
%\end{acknowledgments}

\bibliography{bibliography}

\textbf{Acknowledgements:}
Funding: This study was funded by the Deutsche Forschungsgemeinschaft (DFG, German Research Foundation) - 412351169 within the Emmy Noether program. R.R. acknowledges support from the DFG Walter Benjamin Programme (project RI 3319/1-1). AH acknowledges funding from the European Union’s Horizon 2020 research and innovation programme under the Marie Skłodowska-Curie grant agreement No 841556. I.D.S.  acknowledges support by the Deutsche Forschungsgemeinschaft (DFG, German Research Foundation) under Germany´s Excellence Strategy – EXC 2089/1 – 390776260.
\textbf{Author contributions:}
D.B.B conceived the idea of surface NV-NMR method, designed the experiments and supervised the study. K.S.L. performed all of the surface NV-NMR experiments and was supported by A.H. in ALD depositions and J.D.B. for contact angle measurements. A. H. performed the AFM measurements. M.W.H implemented the phosphonate chemistry and performed XPS measurements. R. D. A. and K.S.L. built the surface NV-NMR set up and imaged the laser spot. R.R. and I.D.S advised on several aspects of theory and experiments.  All authors discussed the results and contributed to the writing of the manuscript. 
\textbf{Competing
interests:}  All authors declare that they have no competing interests. \textbf{Data and
materials availability:} All data needed to evaluate the conclusions in the paper are present
in the paper and/or the Supplementary Materials. Additional data related to this paper may be
requested from the authors. All correspondence and request for materials should be
addressed to D.B.B. (dominik.bucher@tum.de).

%
% ****** End of file apstemplate.tex ******

\clearpage
\onecolumngrid
\begin{center}
\textbf{\large{\textit{Supplementary Information:}{ Surface NMR using quantum sensors in diamonds}}}
\hfill \break

\end{center}

\newcommand{\beginsupplement}{%
        \setcounter{table}{0}
        \renewcommand{\thetable}{S\arabic{table}}%
        \setcounter{figure}{0}
        \renewcommand{\thefigure}{S\arabic{figure}}%
     }
\renewcommand\labelitemi{-}

\beginsupplement
\twocolumngrid
\textbf{Supplementary Note 1: Annealing of implanted diamond for conversion to NV-centers}

\begin{figure}[h]
  \centering
  {\includegraphics[width=0.49\textwidth]{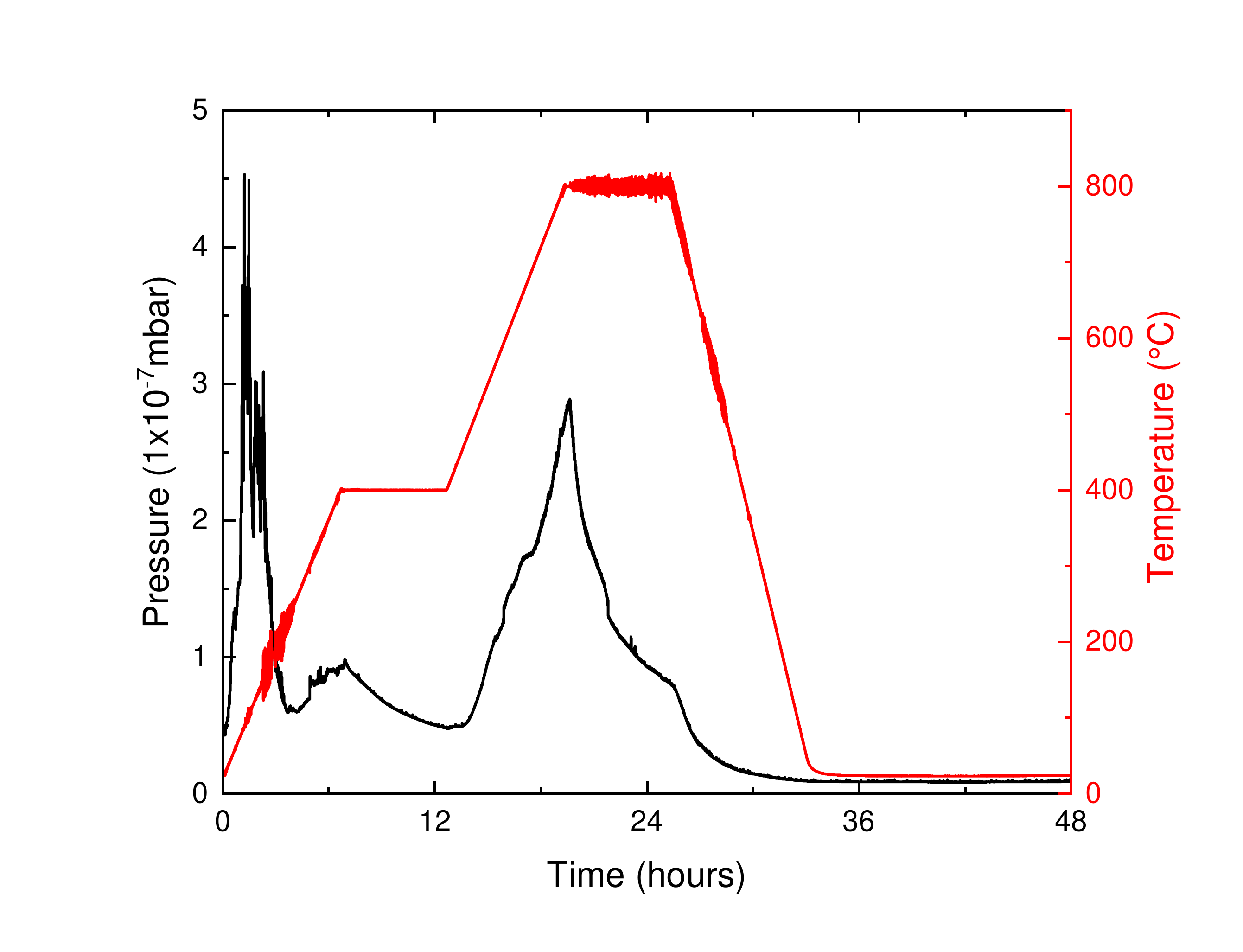}}
  \caption{\textbf{Diamond annealing:} Temperature profile of N-implanted diamond annealing.} 
\label{temperature_profile}
\end{figure}

After implantation, the diamond was annealed in vacuum to form NV-centers using gradual ramp rates and precisely controlled temperatures. The time- and pressure-traces during the annealing process is shown in Fig. \ref{temperature_profile}. The specifications of the oven can be found in the Methods section. The annealing procedure is critical for forming shallow NV-centers, while preserving their sensitivity and brightness and avoiding graphitization and/or damage to the surface \cite{bucherQuantumDiamondSpectrometer2019b}.

\textbf{Supplementary Note 2: Properties of NV-centers after annealing}

\begin{figure}[h]
  \centering
  {\includegraphics[width=0.49\textwidth]{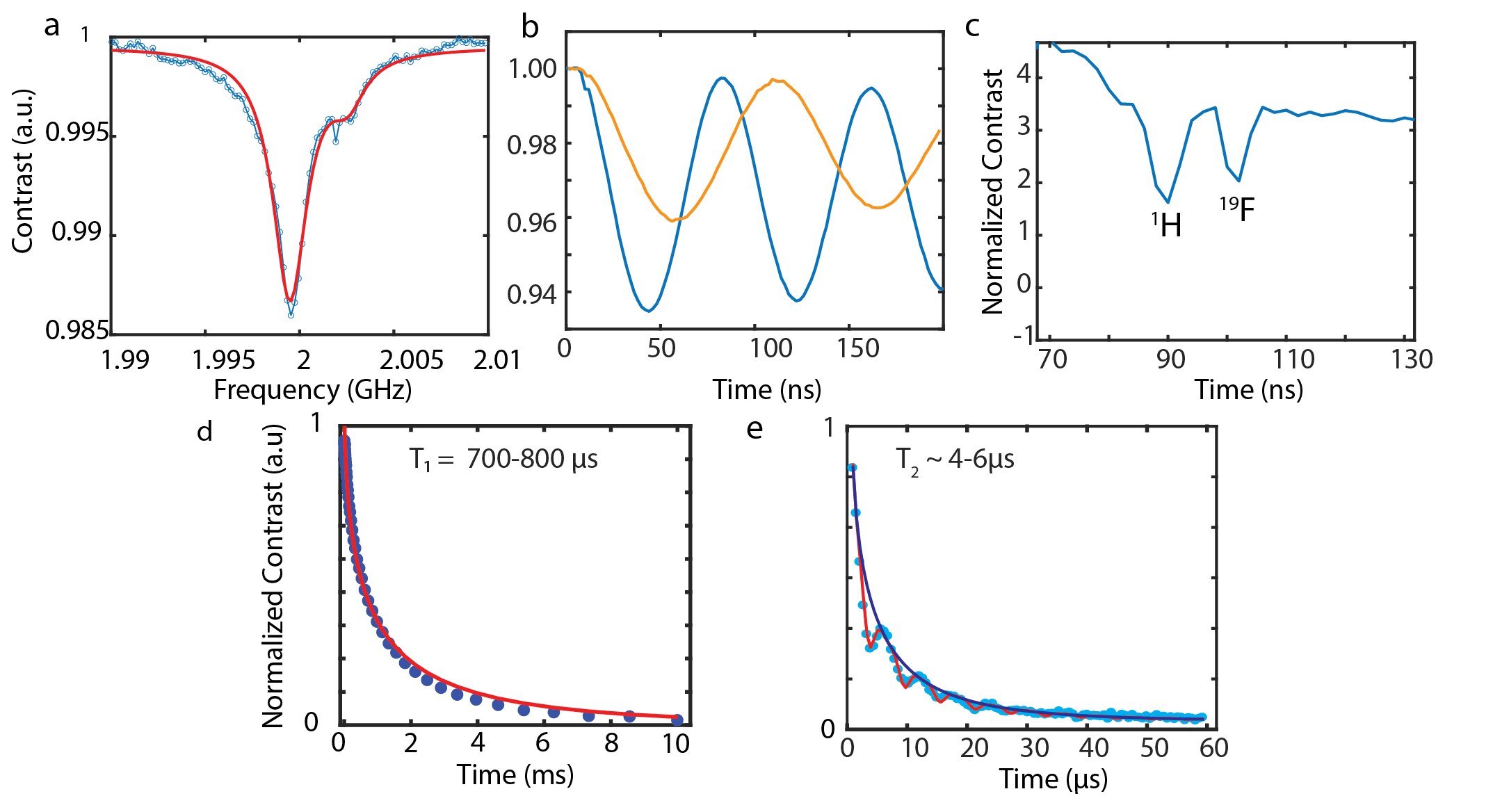}}
  \caption{\textbf{Example of NV ensemble properties } a) ESR lineshape and B$_0$ field b) Typical Rabi contrast dry and in ethanol c) XY8-\textit{12} dips of the $^{1}$H peak on the left and the $^{19}$F peak from drop cast PFOPA on the surface on the right. d) Spin-lattice relaxation T$_1$ E) Spin-spin relaxation T$_2$.}
\label{k_properties}
\end{figure}

The properties of our shallow NV-ensemble in diamond is depicted in Fig. \ref{k_properties}. We used the NV resonance for the m$_{s}$ = 0 to -1 transition to determine the magnetic field strength. The contrast of the Rabi oscillations is typically around 7 $\%$ for dry state (shown in blue), but is reduced to 4 $\%$ in the case of liquid samples on the diamond (shown in yellow). An alternative to the detection of NMR signals on the nanoscale with correlation spectroscopy are dynamic decoupling sequences, where the the time between the $\pi$ pulses is swept \cite{devienceNanoscaleNMRSpectroscopy2015}. Two dips, corresponding to the $^{19}$F and $^{1}$H can be observed in this experiment. The spin-lattice relaxation time T$_1$, is measured by increasing the time between optical initialization and readout. Fitting to an exponential gives a range of 700 to 800 $\mu$s for clean diamond and after ALD. The spin-spin relaxation time T$_2$, is measured with a Hahn Echo sequence. The dips arise from coupling with $^{13}$C in the diamond. The data are fit to a stretched exponential function  resulting in a range from 4-6 $\mu$s for clean diamond and after ALD deposition.

\textbf{Supplementary Note 3: Gaussian fitting of laser spot size}

The laser spot was imaged using a Basler a2A1920-160umBAS camera Fig. (\ref{fig2} c). The entire diamond was imaged with a known size of 2 mm as our reference. Fitting a vertical cut through the center of the laser spot to a Gaussian lineshape allowed for the determination of a full width at half maximum (FWHM) of 60.8 $\mu$m. The horizontal line had a FWHM of 82.9 $\mu$m. Taking these values as the maximal extent of an elliptical area yields a total of $\sim$ 4000 $\mu$m$^2$. 

\begin{figure}[h]
  \centering
  {\includegraphics[width=0.49\textwidth]{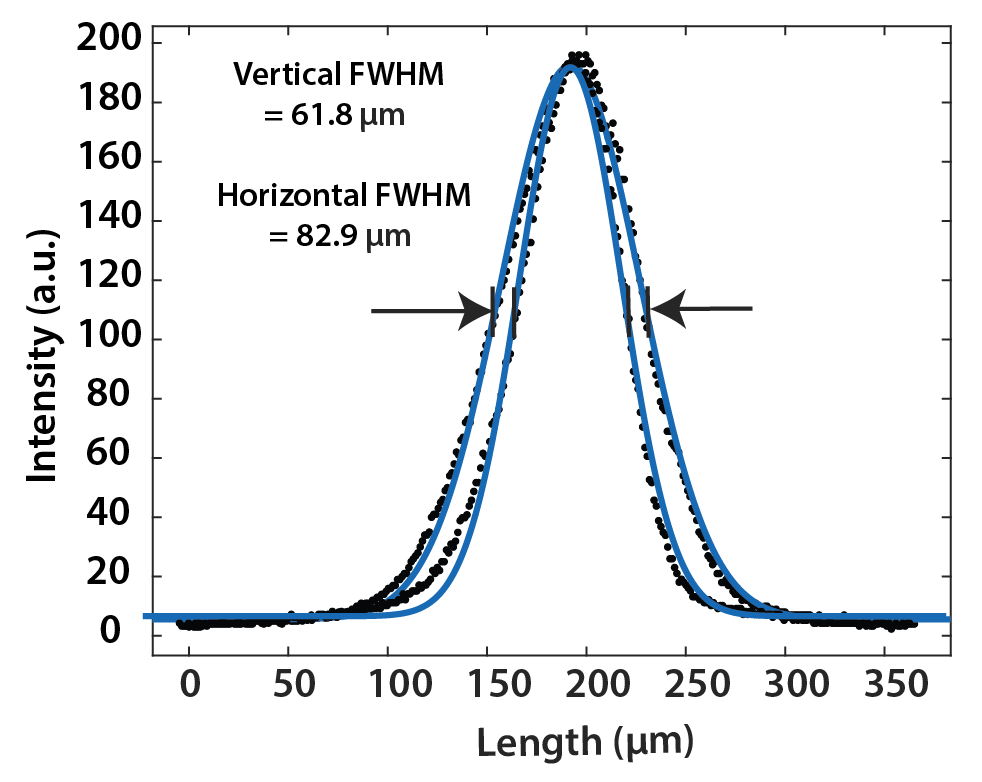}}
  \caption{\textbf{Area of optical excited NV-centers.} The PL spot size of our NV-centers was imaged and the size was determined by vertical and horizontal cuts which were fit to a Gaussian function.}
\label{laser_spot}
\end{figure}

\textbf{Supplementary Note 4: Determining optimal Al$_2$O$_3$ layer thickness for monolayer assembly}

\begin{figure}[h]
  \centering
  {\includegraphics[width=0.49\textwidth]{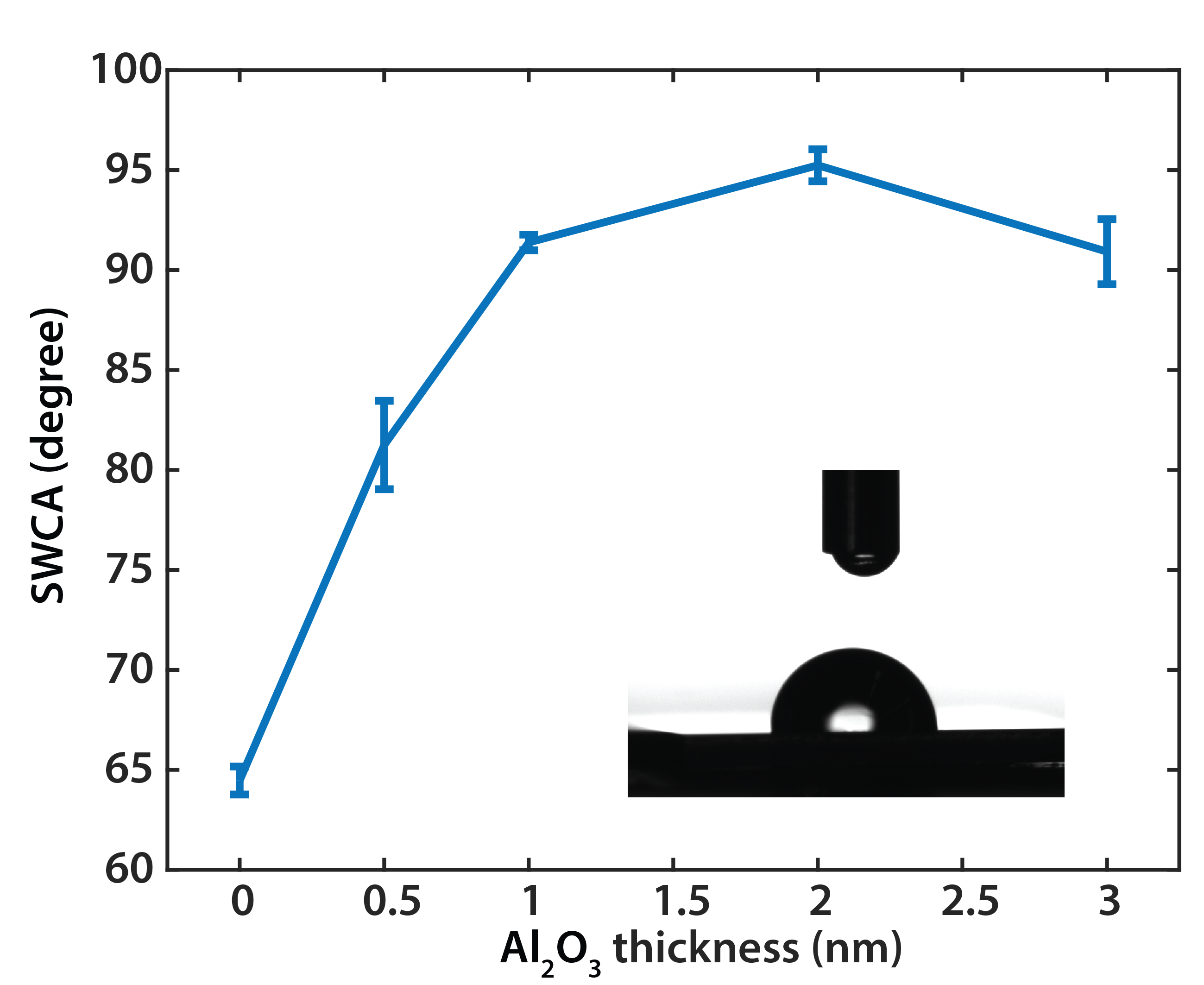}}
  \caption{Static water contact angle as a function of monolayer growth on different Al$_2$O$_3$ layer thicknesses.}
\label{contact_angle}
\end{figure}

As NMR signal decreases with the cubed distance from the NV center, it is crucial to keep the Al$_2$O$_3$ layer as thin as possible. However, it must also be sufficiently thick to form a closed layer that enables assembly of dense molecular monolayers based on phosphonate linking chemistry. We used static water contact angle measurements to determine the thinnest layer of Al$_2$O$_3$ that reaches saturation, as well as to verify dense, ordered monolayer growth from PFPDPA. As a reference, a clean diamond without an Al$_2$O$_3$ layer was soaked in a PA solution without undergoing the ALD process. The low contact angle indicates low reactivity. By contrast, for an Al$_2$O$_3$ layer of 1 nm with a contact angle of approximately 91 degrees the signal indicates that the monolayer reaches a saturation value. Therefore, a thickness of 1 nm was chosen as the thickness of the Al$_2$O$_3$ for surface NV-NMR due to the fact that the nuclear spins are as close as possible to the NV layer for sensing and  thick enough to form a dense monolayer.

%Dense well formed phosphonate monolayers on Al$_2$O$_3$ with long alkyl chains in polar protic solvents have a maximum packing of $\approx$ 5 molecules/ nm$^2$ measured using ellipsometry \cite{dietrichMolecularDynamicsSimulations2017, dietrichMolecularMechanismsSolventControlled2017}. %Our contact angle measurements can be used to determine the density of our monolayer using Cassie's equation $cos(\theta) = f_1(\theta_1) + f_2(\theta_2)$ to relate contact angle to surface coverage, where $\theta$ is the measured contact angle of a surface, $f_2$ the fractional coverage of the surface in question, $\theta_1$ the contact angle for known coverage of a ligand, and $\theta_2$ the contact angle of the bare substrate. A density of 3.3 molecules/ nm$^2$ is calculated with our contact angle measured with PFPDPA monolayer of 90 degrees, 30 degrees measured for Al$_2$O$_3$, and 120 degrees as the typical contact angle for 5 molecules/ nm$^2$ on oxide surfaces and with consideration of tri-dentate binding to the hydroxyl groups at the Al$_2$O$_3$ surface \cite{hotchkissCharacterizationPhosphonicAcid2011, thissenStabilityPhosphonicAcid2010}. This is very reasonable due to our molecular structure and carbon chain length of the phosphonic acid \cite{fukudaEffectsAlkylChain2009,sporiInfluenceAlkylChain2007}. Following this estimation, we find that 16.5 $^{19}$F spins/nm$^2$ and 3.3 $^{31}$P spins/nm$^2$ for our monolayer. 

\textbf{Supplementary Note 5: Magnetic field sweep of $^{31}$P signal}

The $^{31}$P from the PFPDPA monolayer was detected at 3 different magnetic fields. A linear fit of the resonance frequency as a function of magnetic field is used to calculate the gyromagnetic ratio. This corroborates that the source of the signal arises from $^{31}$P spins. The gyromagnetic ratio of 17.24 (17.24, 17.25) MHz/T matches the known value for $^{31}$P.

\begin{figure}[h]
  \centering
  {\includegraphics[width=0.49\textwidth]{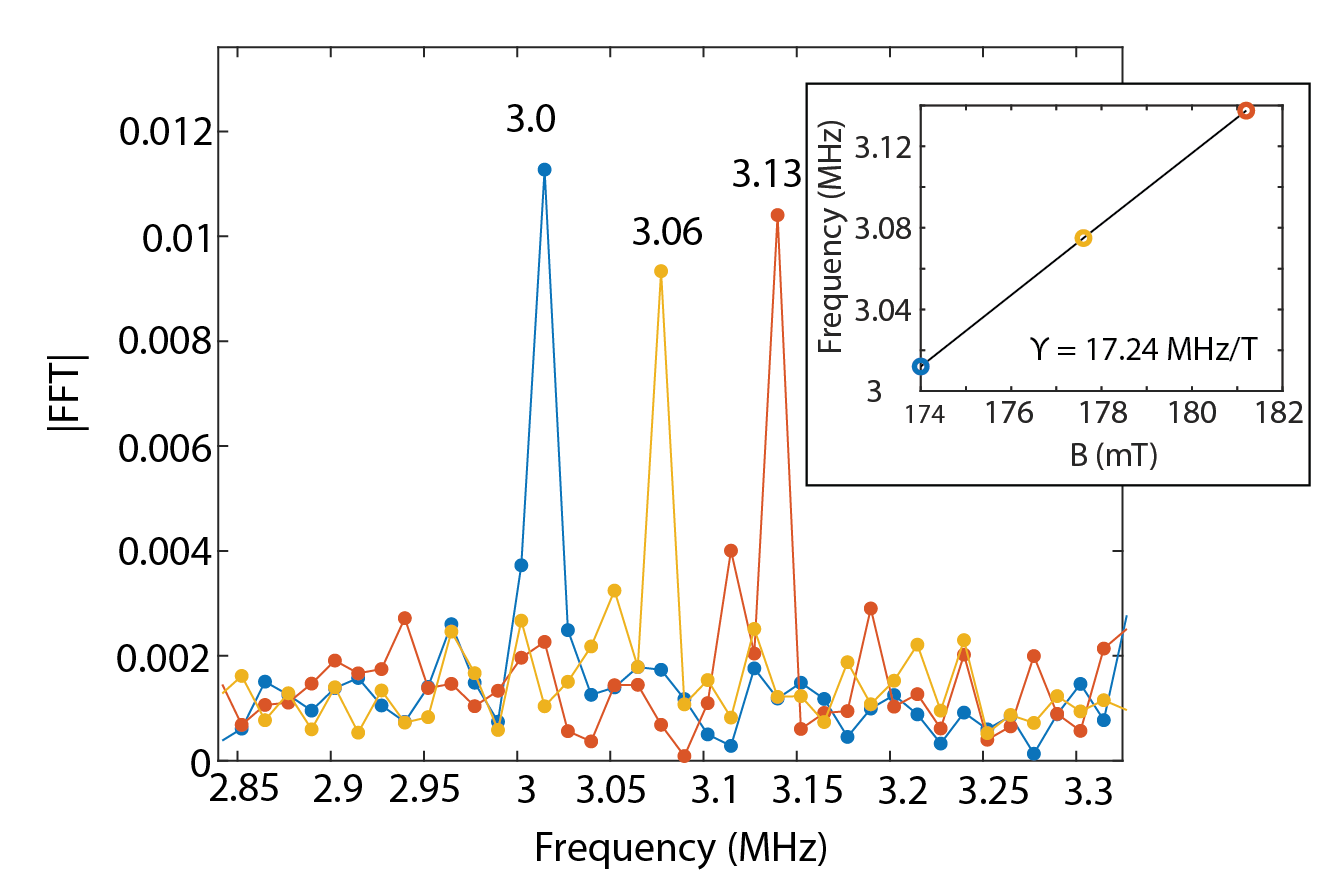}}
  \caption{Magnetic field sweep of $^{31}$P signal from PFPDPA monolayer.}
\label{31P_field}
\end{figure}
\pagebreak

\textbf{Supplementary Note 6: Fitting of \textit{in-situ} growth kinetics experiments}

The rate of monolayer formation can be influenced by changing the concentration of the PFPDPA solution for the \textit{in-situ} growth kinetics experiments. The averaged data points for each concentration were fit with a single exponential: $y= a(1-x^{-b})+c$. We observe an increase in the rate with an increase in concentration (Table S1).

\begin{table}[h]
 \caption{Fitting parameters of the monolayer growth kinetics}
    \begin{tabular}{|c|c|c|c|c|}
    \hline
     [Conc.] & $a$ & $b$ & $c$ \\ 
    \hline
     1 mM &  0.81 & 0.10 & 0.13 \\ 
    \hline
    10 $\mu$M  & 0.084 & 0.052 & 0.13 \\ 
    \hline
     1 $\mu$M & 0.093 & 0.039 & 0.027 \\ 
    \hline
    \end{tabular}
\end{table}
\pagebreak

\textbf{Supplementary Note 7: Diamond surface cleaning for redeposition of Al$_2$O$_3$}

\begin{figure}[h]
  \centering
  {\includegraphics[width=0.49\textwidth]{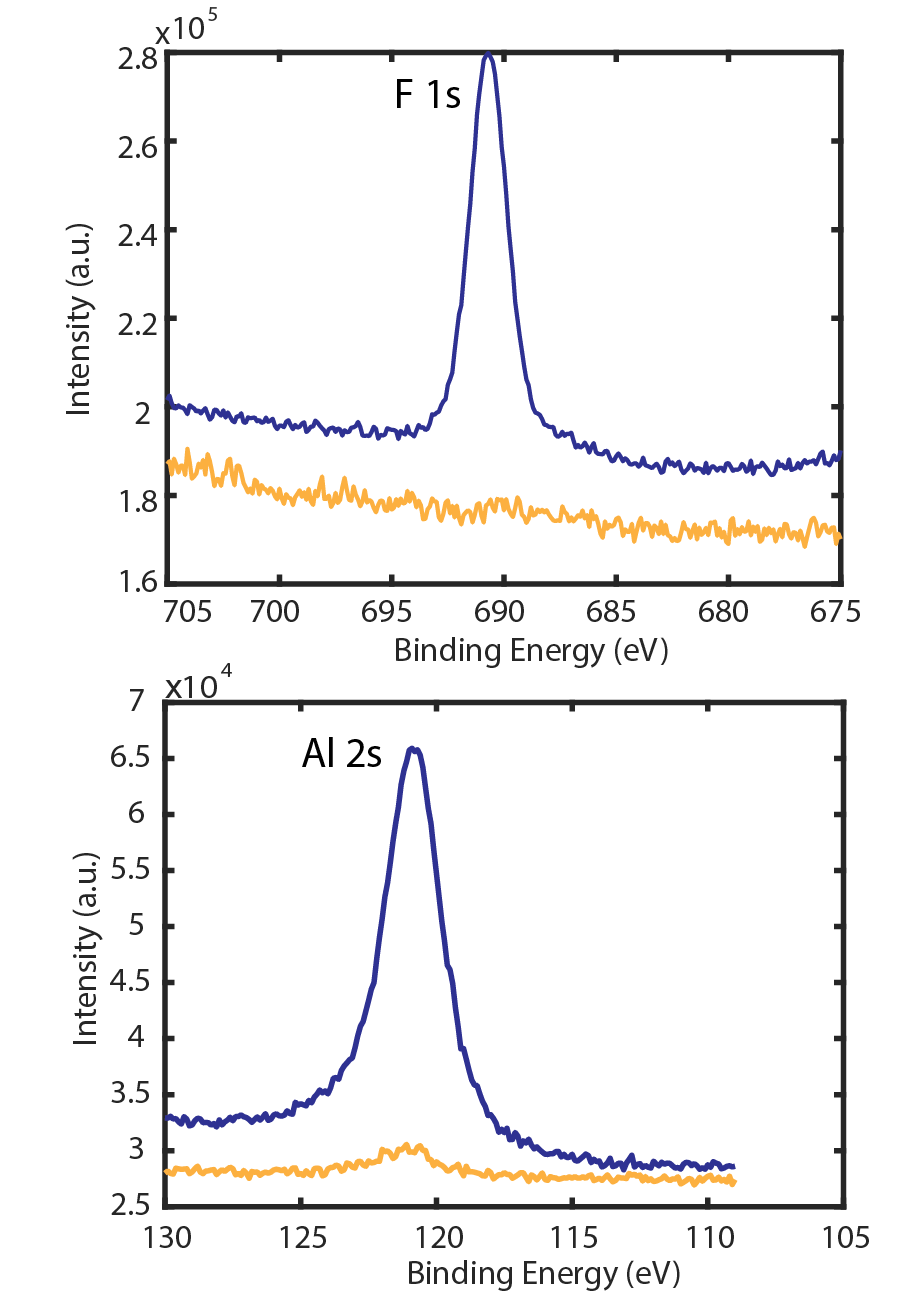}}
  \caption{XPS of the surface before and after Al$_2$O$_3$ removal with NaOH} 
\label{NaOH_removal}
\end{figure}

To be able to reproduce the ALD and monolayer growth process, removal of the Al$_2$O$_3$ is crucial. XPS measurements of F 1s and Al 2s peaks before (blue) and after (yellow) overnight soaking in 5$\%$ NaOH shows that NaOH is effective at completely removing the monolayer and Al$_2$O$_3$.

\textbf{Supplementary Note 8: Materials and Methods}

\textbf{Surface NV-NMR setup.} The diamond was positioned in the middle of two neodymium magnets, which were rotated and tilted for alignment of the B$_0$ field with one of the four possible NV center orientations. For quantum control of the NV-centers, microwave frequencies were generated with a signal source (SynthHD, Windfreak Technologies, LLC., New Port Richey) and fed into a phase shifter (ZX10Q-2-27-S+, Mini-Circuits) and two switches (ZASWA-2-50dRA+, Mini-Circuits) to generate X and Y pulses, then combined (ZX10-2-442-s+) and amplified by a microwave amplifier (ZHL-16W-72+, Mini-Circuits). Initialization of the NV ensemble was achieved by using a 532 nm laser (Verdi G5, Coherent) at a power of around 200 mW at the diamond. The laser pulses were controlled by an acousto-optic modulator (Gooch and Housego, model 3260-220) with pulse durations of 5 $\mu$s. The diamond was glued to a thin glass slide (48393-026, VWR). The photoluminescence (PL) was collected and collimated by 2 condenser lenses, the bottom one placed right above a large area avalanche photodiode (A-CUBE-S3000-10, Laser Components GmbH) below the diamond. In order to increase the light collection efficiency, a 6 mm glass hemisphere (TECHSPEC\textsuperscript{\textregistered} N-BK7 Half-Ball Lenses, Edmund Optics) was glued to the other side of the cover slide. This assembly was then glued to a 30 mm cage plate (CP4S, Thorlabs) and mounted onto the experiment 1.2 cm above the top condenser lens (ACL25416U-B, Thorlabs). This coupled in the focused laser (LA1986-A-M, Thorlabs) in a total internal reflection geometry. The excitation wavelength was removed from the PL light with a long-pass filter (Edge Basic 647 Long Wave Pass, Semrock) placed immediately between the bottom condenser lens and the photodiode. The photo-voltage was digitised with a data acquisition unit (USB-6229 DAQ, National Instruments). The electron spin resonance (ESR) frequency measured from the dip in PL was used to determine the magnetic field strength and the NV resonance frequency to perform a Rabi experiment, which then determined the $\pi$ and $\pi$/2 pulse durations for the correlation spectroscopy pulse sequences Fig. (\ref{k_properties}). The magnetic field strength B$_0$ can be adjusted by changing the distance between the magnets. We chose to work at 31 mT for $^{19}$F and at 174 mT for $^{31}$P detection. The lower magnetic field was suited for higher frequency detection because the spacing between the microwave pulses must be 1/(2$\tau$) which was limited by finite $\pi$ pulse durations.

\textbf{SWCA measurements.} SWCA measurements were performed on an OCA 15Pro contact angle system (DataPhysics Instruments). Data acquisition and evaluation were realized with SWCA 20 - contact angle (DataPhysics Instruments, version 2.0). For quantifying an average Young’s contact angle ($\theta\gamma$), 2 $\mu$L of deionized H$_2$O (18.2 M $\Omega$cm at 25 $^{\circ}$C, Merck Millipore) was dispensed with a rate of 0.2 $\mu$Ls$^{-1}$ from a 500 $\mu$L Hamilton syringe onto the sample surface. After allowing the droplet to settle for $\sim$ 3 s, an image was acquired for further processing. The procedure was repeated at least 3
times on different spots on the surface, and the standard deviation (error) was calculated.

\textbf{Atomic Force Microscopy.} Bruker MultiMode 8, (Bruker Corp.) was used in tapping mode and in contact mode under ambient conditions using NSG30 (TipsNano) for standard characterization and to estimate the Al$_2$O$_3$ layer thickness. Scratching in contact mode was performed over areas of 1 x 1 $\mu$m$^2$ with a deflection set point of 5 V. Amplitude modulation (tapping mode) AFM was done with an amplitude set point of 0.3 V (at a free amplitude of 0.5 V). The surface roughness was evaluated via the RMS average of height deviations taken from the mean image data plane of 2 x 2 $\mu$m$^2$ tapping-mode micrographs. The roughness and step heights were analyzed using Gwyddion 2.56.

\textbf{X-ray Photoelectron Spectroscopy.} X-ray photoelectron spectroscopy (XPS) measurements were performed with an Axis Supra (Kratos, UK) spectrometer. The monochromatized Al K$_\alpha$ (1486.9 eV) X-ray tube source was operated at an emission current of 15 mA. The data were recorded with a circular acceptance area of Al-source and the analysed area 700 x 300 $\mu$m in diameter. Spectra were processed with CasaXPS (Casa Software Ltd,
version 2.3.17).
\end{document}